\documentclass[12pt,notoc]{JHEP3}
\usepackage[fleqn]{amsmath}
\usepackage[latin1]{inputenc}
\usepackage{amssymb,bbm}

\def\del{\partial}
\def\e{\mathrm{e}}
\def\abs#1{\left|#1\right|}
\def\dilaton{{\mathit{\phi}}}
\def\phn{{\phantom{0}}}
\renewcommand*{\i}{\mathrm{i}}
\newcommand*{\ie}{i.\,e.\ }

\newcommand*{\cf}{cf.\ }
\renewcommand*{\d}{\mathrm{d}}
\newcommand*{\dx}{\d x}
\newcommand*{\dz}{\d z}

\newcommand*{\dt}{\d t}
\newcommand*{\ds}{\d s}
\renewcommand* {\vec}[1]{\mathbf{#1}}

\title{
  SL(2,Z) tensionless string backgrounds in IIB string theory
}

\author{
  Andreas Bredthauer, Ulf Lindstr{\"o}m, Jonas Persson\\
  Department of Theoretical Physics\\
  Uppsala University\\
  Box 803, SE-751\,08 Uppsala\\
  {\tt Andreas.Bredthauer, Ulf.Lindstrom, Jonas.Persson@teorfys.uu.se}
}
\abstract{
We examine a tensionless limit of a SL(2,Z) set of background solutions
to IIB supergravity theory, obtained by performing an infinite 
boost. This yields a solution that corresponds to taking the 
original string tension to zero. The limit reproduces ordinary Minkowski 
space except for a delta-like singularity along the string. We study the 
field content and the energy momentum tensor.
}

\preprint{
  hep-th/0303225\\
  UUITP-04/03
}


\begin{document}

\section{Introduction}

In string theory the analogue of a massless particle is a tensionless string
\cite{Schild:vq,Karlhede:wb}. Just like particles often can be treated as 
approximately massless in a high energy description, tensionless strings are 
expected to be of interest for the high energy behavior of string theory 
\cite{Gross:1987kz,Gross:sj}. Various action formulations with or without 
supersymmetry have been constructed 
\cite{Barcelos-Neto:1989ms,Zheltukhin:1989kx,Lindstrom:1990qb,
Lindstrom:1990ar,Zheltukhin:2001jw}, and their quantum properties have been 
studied \cite{Lizzi:1986nv,Isberg:1992ia,Isberg:1993av,Gustafsson:1994kr,
Saltsidis:1995qr}. One interesting result is that if the (spacetime) 
conformal symmetry of the classical formulation is to survive quantization, 
the spectrum is topological \cite{Gustafsson:1994kr}. (This brings to mind 
the comments by Witten on the relation of the results in 
\cite{Gross:1987kz,Gross:sj} to topological theories \cite{Witten:1988zd}).

It should also be noted that in the AdS/CFT context, an interesting relation 
between the spectrum of tensionless strings and higher spin theories has been 
advocated \cite{Sundborg:2000wp}.\footnote{\cf \cite{Vasiliev:1999ba} and 
references therein for an overview on highes spin theories.}
More abstractly, tensionless strings have recently been seen to arise in the 
framework of intersecting M5-branes \cite{Constable:2002vt} and earlier in 
the context of D-branes \cite{Witten:1995zh,Sen:1996tz,Strominger:1996ac,
Hanany:1996hq,Ganor:1996mu}, although for these tensionless strings an action
is lacking.

An important question is the role of these strings in the solution space of 
supergravity. In this brief note we show that they arise as sources to exact 
solutions therein. It proceeds as follows: First, we describe 
Schwarz' classical solution to $D=10$ IIB supergravity theory 
\cite{Schwarz:1995dk} in terms of a set of $SL(2,{\mathbbm Z})$ background 
metrics, which is the starting point for our calculation. Then, in section 3, 
we perform a Lorentz transformation orthogonally to the string. We derive the 
limit as the boost achieves the speed of light. It turns out that the 
solution is Minkowski space up to a plane-fronted shock wave at the position 
of the (boosted) string. We show that this gives a solution to the string 
equations to all orders and that the limit is compatible with Einstein's 
equations. We derive the energy momentum tensor and show that is has the 
expected behavior. We end with a few comments and open questions for future 
study. 

\section{The original solution}

We consider $D=10$ IIB supergravity theory with the bosonic action
\begin{align}
  S^{\mathrm{IIB}}_{10}=\frac{1}{2\kappa^2}\int\d^{10}x\sqrt{-g}\left[
  R+\frac{1}{4}\mbox{tr}(\partial {\cal M}\partial {\cal M}^{-1})-
  \frac{1}{12}{\cal H}^T{\cal M}{\cal H}\right], \label{eqn:theaction}
\end{align}
where $F_5$ is put to zero.\footnote{The charges of the $F_5$ field 
strength are carried by 3-branes, which is not of interest here.} ${\cal M}$ 
is a $SL(2,{\mathbbm R})$
matrix of the scalar fields and ${\cal H}$ is a vector of the 3-forms:
\begin{align}
  {\cal M}=\e^\dilaton\left(\begin{array}{cc}
    \abs{\lambda}^2&\chi\\
    \chi&1
   \end{array}\right),\hspace*{1cm}
  {\cal H}=\d{\cal B}=\left(H^{(1)},H^{(2)}\right),
\end{align}
with $\lambda=\chi+\i\e^\dilaton$. Here, $\dilaton$ is the dilaton (NS-NS) 
field and $\chi$ is the R-R scalar field. The action \eqref{eqn:theaction} 
is invariant under the following action of $SL(2,{\mathbbm R})$:
\begin{align}
  {\cal M}\rightarrow\Lambda{\cal M}\Lambda^T,\hspace*{1cm}
  {\cal B}\rightarrow(\Lambda^T)^{-1}{\cal B}.
\end{align}
Schwarz generalized the results of Dabholkar et al.\ \cite{Dabholkar:1990yf} 
and found a solution given by the $SL(2,{\mathbbm Z})$ set of background 
metrics \cite{Schwarz:1995dk}:
\begin{align}
\label{originalmetric}
\ds^2 = A_q^{-3/4}\left(-\dt^{\,2} + (\dx^1)^2 \right)
        + A_q^{1/4}\,\dx\cdot\dx.
\end{align}
where $\dx = \left(\dx^2, \ldots, \dx^9\right)$,
\begin{align}
A_q = 1 + \frac{\Delta_q^{1/2} Q}{3r^6},\hspace*{1cm}
\Delta_q^{1/2} = \vec{q}^T{\cal M}^{-1}\vec{q},~~\vec{q}=(q_1,q_2).
\end{align}
Herein, $r^2 = x \cdot x$ and $Q$ is the fundamental ${\cal B}_{\mu\nu}$ 
electric charge. The restriction to $SL(2,{\mathbbm Z})$ follows 
from the Dirac quantization condition and that $q_1$ and $q_2$ are 
relative prime. The corresponding fields read
\begin{align}\label{Bequation}
&{\cal B}_{01} = {\cal M}^{-1}\vec{q}\, \Delta_q^{-1/2} A_q^{-1},\\
&\lambda = \frac{q_1\chi_0- q_2 |\lambda_0|^2 + \i q_1 \e^{-\phi_0}A_q^{1/2}}
                 {q_1 - q_2\chi_0 +\i q_2\e^{-\phi_0}A_q^{1/2}},
\end{align}
where we used the fact that the only non-zero components of ${\cal B}$ can 
be taken to be ${\cal B}_{01}= - {\cal B}_{10}$ \cite{Dabholkar:1990yf}. 
The metric has a singularity at $r=0$, which is interpreted as an infinitely 
long source string with action \cite{Schwarz:1995dk, deAlwis:1996ze}
\begin{align}\label{sourceaction}
S = -\frac{T_q}{2}\int \d^2\xi \left[\Delta^{1/2}_q
     \partial^a X^\mu \partial_a X^\nu G_{\mu\nu}
     + \epsilon^{ab}\partial_a X^\mu \partial_b X^{\nu}
     {\cal B}^T_{\mu\nu}\vec{q}\right].
\end{align}
Here, $T_q=\Delta^{1/2}_q Q$ is the string tension. 
The background fields in \eqref{sourceaction} are actually string condensates.
They arise as string loop effects \cite{deAlwis:1996ze}. 
The above string solutions were interpreted by Witten as bound states of 
F(undamental) and D-strings \cite{Witten:1996im}.

\section{Performing the boost}
We study a boost of the metric \eqref{originalmetric} in a direction 
orthogonal to the string. Without loss of generality, we perform a Lorentz 
transformation in the $z=x^9$ direction.  The form of the transformation is 
the usual
\begin{align}\label{lorentztransf}
  t'= \gamma (t+vz),~~ 
  z'=\gamma(x+vt),~~ \gamma=\left(1-v^2\right)^{-1/2}.
\end{align}
In the transformed coordinates, the metric reads
\begin{multline}\label{metric}
\ds' = \frac{A_q^{\prime -3/2}\Delta_q^{1/2} Q \gamma^2}{3r^{\prime 6}}
         \left(\dt'-v\dz'\right)^2
       +A_q^{\prime -3/4}(\dx^1)^2
       \\+A_q^{\prime 1/4}\left(-(\dt')^2 + \d\tilde{x}\cdot\d\tilde{x} 
           + (\dz')^2\right),
\end{multline}
where we introduced $\tilde{x}= \left(x^2, \ldots x^8\right)$,
$r^{\prime 6} = \left(\gamma^2(z'-vt')^2 + \tilde{x}\cdot\tilde{x}\right)^3$ 
and 
\begin{align}
A'_q = 1 + \frac{\Delta_q^{1/2} Q}{3r^{\prime 6}}.
\end{align}
If we boost \eqref{metric} to the speed of light, \ie let $v\rightarrow 1$, 
we immediately find that the energy diverges. To avoid this problem, the 
energy has to be rescaled as $\epsilon = \gamma^{-1}\epsilon_0$ or, 
equivalently, $Q=\gamma^{-1}Q_0$, following the lines of Aichelburg and Sexl 
\cite{Aichelburg:1971dh}. We notice that $\lambda$, ${\cal M}$ and hence, 
$\Delta_q$ tend to their vacuum expectation values $\lambda_0$, ${\cal M}_0$
and $\Delta_{q,0}$, respectively. This implies, in particular, that 
the dilaton field $\phi$ and the R-R scalar field $\chi$ tend to their 
(constant) vacuum expectation values. The string tension becomes
$T_q = \Delta_q^{1/2}Q \rightarrow \gamma^{-1}\Delta_{q,0}^{1/2}Q_0^\phn$ and 
vanishes as $v\rightarrow 1$. Thus, the result may be interpreted as a 
background metric with a tensionless string as its source. Finally, we find
$A_q^\prime\rightarrow 1$ in the boost limit. Therefore, the only term we 
need to consider more carefully is the first term in \eqref{metric}, \ie 
the limit of 
\begin{align}
  \gamma r^{\prime -6} = \frac{\gamma^{-5}}{\left(\left(z'-vt'\right)^2 + 
  \gamma^{-2}\rho^2\right)^3}, \label{rprime}
\end{align}
where $\rho^2 = \tilde{x}\cdot\tilde{x}$. Integrating both sides, one finds 
that the right hand side of \eqref{rprime} tends to
\begin{align}
  \lim_{v\rightarrow 1} \gamma r^{\prime -6}=\frac{3\pi}{8} 
       \left(\tilde{x}\cdot\tilde{x}\right)^{-5/2} 
       \delta\left(z'-t'\right),
\end{align}
Thus, in the boost limit, the metric becomes
\begin{align}
\ds' = \frac{\pi\Delta_{q,0}^{1/2} Q_0^\phn}{8\rho^5} \delta\left(z'-t'\right) 
       \left(\dt'-\dz'\right)^2 +(\dx^1)^2 
       -(\dt')^2 + (\dz')^2 + \d \tilde{x}\cdot\d \tilde{x},
\end{align}
or, equivalently, in coordinates $u=z'-t',~v=z'+t'$,
\begin{align}\label{boostedmetric}
(\ds')^2 = \d u \d v + (\dx^1)^2 + \d\rho^2 + \rho^2\d\Omega 
          + \frac{\pi\Delta_{q,0}^{1/2} Q_0^\phn}{8\rho^5} 
            \delta\left(u\right) \d u^2.
\end{align}
The solution is still an $SL(2,{\mathbbm Z})$ set of background
metrics. Moreover, it has the typical pp-wave metric structure
\begin{align}
  \ds^2 = \d u \d v + K(x^2, \ldots, x^8, u)\d u^2 + \sum_{i=1}^8 (\d x^i)^2.
\end{align}
For boosts in other directions we find an expected factor of $\sin^2 \phi$
in front of the coefficient function $K$, where $\phi$ is the angle between
the boost direction and the direction of the string, \ie $x^1$. Especially,
the only direction significantly different from the one described above 
($\phi=\pi/2$) is the direction parallel to the string ($\phi=0$) where we 
recover flat Minkowski space and in which case the $SL(2, {\mathbbm Z})$
degeneracy vanishes.

\subsection{The B field}

According to \eqref{Bequation}, we only consider the transformation of
${\cal B}_{01}$. The Lorentz transformation \eqref{lorentztransf} 
generates four non-zero components in the transformed ${\cal B}'$ field, 
all of which diverge to infinity in the limit $v\rightarrow 1$. 
This apparent problem is resolved by using the gauge freedom to perform
the transformation
\begin{align}
{\cal B}_{01}\rightarrow \tilde{\cal B}_{01} 
   = {\cal B}_{01} - {\cal M}^{-1} \vec{q}\, \Delta_q^{-1/2}
   = {\cal M}^{-1} \vec{q}\,\Delta_q^{1/2} \left(A_q^{-1} - 1 \right) 
   \stackrel{v\rightarrow 1}{\longrightarrow} 0.
\end{align}
The gauge fixed ${\cal B}$ field vanishes in the limit of the boost, 
and hence does ${\cal H}=\d {\cal B}$.

\subsection{Energy momentum tensor}

The boosted metric $G_{\mu\nu}$ given by \eqref{boostedmetric} is "mostly 
flat" and contains a plane fronted shock wave at $u=0$, \ie along the 
position of the string, encoded in the coefficient function $K$. This 
implies that $G_{\mu\nu}$ solves Einstein's equations in empty space except 
for $\rho=0$ which contains a coordinate singularity. Moreover, since 
$1/\rho^5$ is the Green's function for the Laplacian in seven dimensions, 
\ie
\begin{align}
  \Delta\frac{1}{\rho^5}=-\frac{16}{3}\pi^3\delta(\rho),
\end{align}
we obtain
\begin{align}
  R_{uu}=-\frac{1}{2}\Delta K=
    \frac{1}{3}\pi^4\Delta^{1/2}_{q,0} Q_0^\phn\delta(\rho)\delta(u).
\end{align}
All other components of the Ricci tensor vanish. Thus, since the Ricci 
scalar vanishes, the energy momentum tensor of the system is read off from
Einstein's equations:
\begin{align}\label{eqn:Tmunu}
  T_{uu}=\frac{1}{24}\pi^3\Delta^{1/2}_{q,0} Q_0^\phn\delta(\rho)\delta(u).
\end{align}
This solution has the expected form for an energy momentum tensor
of an infinitely long (classic) string at rest. However, the result has to be 
compared to the solution directly derived from the action. The energy 
momentum tensor for non-vanishing tension is proportional to $T_q$. In the 
case of vanishing tension, it can be obtained from the action
\begin{align}
  S=-\int \d^2\sigma 
    \Lambda\left(\del_a X^\mu\del^a X^\nu G_{\mu\nu}
    \right)^2. 
\end{align}
This is the original source term \eqref{sourceaction} written with the help of 
a Lagrange multiplier and having performed the limit, \ie 
$T_q, {\cal B}_{\mu\nu}\rightarrow 0$. A better approach is to rewrite the
action with the help of a world-sheet vector density $V^a$ of weight 
$\frac{1}{2}$ \cite{Lindstrom:1990qb,Lindstrom:1990ar}
\begin{align}
  S=\int \d^2\sigma V^a V^b \del_a X^\mu\del_b X^\nu G_{\mu\nu}.
\end{align}
The tensor quantity $V^a V^b$ has determinant zero and $V^a$ satisfies the 
following equations of motion:
\begin{align}
  V^b\del_a X^\mu\del_b X^\nu G_{\mu\nu}=0, \hspace*{1cm}
  \del_a\left(V^a V^b \del_b X^\mu\right)=0.
\end{align}
For simplicity, we restrict ourselves to the transverse gauge on the 
world-sheet by using the reparametrization symmetry, \ie fixing
\begin{align}
  V^a=\left(\begin{array}{c}v\\0\end{array}\right).
\end{align}
Moreover, by using the diffeomorphisms on the world-sheet, we fix the 
constant $v$ to $v=1$. This is, in fact, the so called diffeomorphism gauge,
yielding
\begin{align}
  T_{\mu\nu}(x^\sigma)=
    \del_\tau X_\mu\del_\tau X_\nu\delta^{(8)}(X^\sigma-x^\sigma),
\end{align}
where the eight-dimensional delta function covers the space transverse to the 
string, \cf \cite{Gurses:1975cm}. Here, we already integrated out the 
world-sheet directions. This implies that $X^u$ and $X^1$ are fixed to the 
values of $x^u$ and $x^1$. Since $X^2, \ldots, X^8, X^u=0$, the only 
non-vanishing contribution arises from 
\begin{align}
  X_u=G_{uv}X^v=X^v=\alpha\cdot \tau.
\end{align}
From this, we obtain
\begin{align}
  T_{uu}=\del_\tau X_u\del_\tau X_u\delta(\rho)\delta(u)=
  \alpha^2\delta(\rho)\delta(u),
\end{align}
and thus, by comparing to \eqref{eqn:Tmunu},
\begin{align}\label{eqn:alpha}
  \alpha^2=\frac{1}{24}\pi^3\Delta^{1/2}_{q,0} Q_0^\phn.
\end{align}
Therefore, the momentum density of the (boosted) string is fixed by 
\eqref{eqn:alpha}. The more important outcome is that \eqref{boostedmetric} is 
"compatible" with taking the limit directly inside the action.

\section{Conclusions}

We found an $SL(2, \mathbbm{Z})$ family of string backgrounds describing 
tensionless strings by performing an infinite boost to the solution of 
Schwarz \cite{Schwarz:1995dk}. The outcome is of pp-wave form incorporating
a delta-like singularity along the position of the string. We showed that
the anti-symmetric tensor field ${\cal B}$ vanishes in this limit and hence
do the 3-forms. The dilaton as well as the R-R scalar field become
constant. We calculated the energy momentum tensor and found that is
has the expected properties.

Hayashi and Samura performed a Penrose limit \cite{Blau,Blau:2001ne} on 
solutions of the type we found \cite{Hayashi:1996qx}. Considerations 
similar to our were done by Horne et.\,al.\ \cite{Horne:1992cn}. However,
their solution did not contain a delta-like singularity and this was believed
to be true for any boosted string.
 
As there is not much known about tensionless strings themselves, it is an 
interesting open question for further studies to look at the spectrum that 
arises from this new solution. On the other hand, tensionless strings arise 
naturally in the context of intersecting branes. Consequently, one might 
think of relating the presented results to tensionless strings arising in 
M-theory. One could also envisage connecting this to the context of string 
scattering from D-branes, \cf Garousi and Myers \cite{Garousi:1996ad}.

\acknowledgments
UL acknowledges initial discussions with Bo Sundborg on the topic of this 
article. The authors further gratefully acknowledge discussions with James 
Gregory and Chris Hull. The research of UL is supported in part by EU 
contract HPNR-CT-2000-0122 and by VR grant 650-1998368.

\providecommand{\href}[2]{#2}\begingroup\raggedright\endgroup

\end{document}